# Frequency-swept high-repetition rate optical source

**Christophe Finot**

Laboratoire Interdisciplinaire CARNOT de Bourgogne,
UMR 6303 CNRS-Université de Bourgogne-Franche-Comté,
9 Av. A. Savary, BP 47 870, 21078 DIJON Cedex, FRANCE

christophe.finot@u-bourgogne.fr

*Abstract*

We propose and numerically validate an all-optical scheme to generate optical pulse trains with varying temporal pulse-to-pulse delay and pulse duration. Applying a temporal sinusoidal phase modulation followed by a shaping of the spectral phase enables us to maintain high-quality Gaussian temporal profiles.

*Keywords:* Electro-optic modulators, optical pulses, microwave photonics, phase modulation, pulse width modulation.

## 1. Introduction

The modulation of the temporal optical phase by a sinusoidal waveform is a simple but very efficient method to all-optically process the properties of a signal at very high-repetition rates. Various examples have indeed been demonstrated in the context of fiber ultrafast optics such as optical time lens [1], mitigation of self-phase modulation [2], improvement of nonlinear spectral compression process [3], enhancement of the extinction ratio [4], to cite a few … Another field of application that has stimulated strong interest is the photonic generation of optical pulse trains at repetition rates of several tens of GHz. Indeed, applying a sinusoidal temporal phase modulation on a continuous wave followed by a spectral phase shaping has been found to be a promising approach that can overcome the bandwidth limitations of the usual optoelectronics modulation schemes. Initial developments were based on a quadratic spectral phase to convert the temporal phase modulation into a modulation of the intensity profile [5, 6]. However, this scheme suffers from a detrimental residual background that impairs the picosecond pulse train quality [7]. The method has been recently improved by imprinting a triangular spectral phase profile or equivalently, a series of equally spaced $\pi/2$ phase shifts. High-quality pulse trains made of Fourier transform-limited Gaussian pulses have been experimentally demonstrated [8]. This architecture can compete with alternate cavity-free approaches based on Kerr nonlinear elements [9, 10] and has the advantage to sustain multi-wavelength operation [11]. It has also been shown that starting from a two-tone sinusoidal modulation enables the generation of a pulse train with controllable levels of fluctuations of the peak power and temporal duration [12].

In this contribution, we further extend this approach by investigating the possibility of delivering pulse trains with a continuously varying pulse-to-pulse delay. By sweeping the frequency of the sinusoidal modulation, we numerically show that we can tune the properties of the pulses while maintaining an excellent pulse quality, both in terms of duty cycle, extinction ratio and waveform. The scheme that is chosen to process the spectral phase is crucial and very different behaviors can be observed.

We first introduce the principle of our approach and the different configurations we have studied and compared. Results of numerical simulations are then discussed.

## 2. Principle of operation

Our method is based on the principle detailed in [8]. A continuous optical wave with an amplitude $\psi_0$ and a carrier frequency $f_c$ is temporally phase modulated by a sinusoidal waveform $A_m \cos(2\pi f_m(t)\, t)$ with $A_m$ being the amplitude of the phase modulation (chosen as $A_m = 1.1$ rad following the guidelines of [8]) and $f_m(t)$ its non-constant frequency. In the present work, we consider a linear temporal sweep of the frequency of the RF signal driving the modulator between $f_{m1}$ and $f_{m2}$ so that $f_m(t) = f_{m1} + (f_{m2} - f_{m1})\, t\, /\, T_s$, $2\, T_s$ being the time of the frequency sweep. In order to illustrate our discussion with numerical simulations based on realistic parameters, we have chosen the boundaries $f_{m1}$ and $f_{m2}$ being 30 and 35 GHz respectively. $T_s$ is fixed by the RF generator and does not impact the results under discussion. The optical spectrum obtained after the temporal phase modulation is plotted in Fig. 1(a) and is symmetric.

Compared to the frequency-constant case modulation where the spectrum is made of equally-spaced spectral lines with an amplitude provided by Bessel functions of the first kind [13, 14], the energy of the spectral components is here spread on spectral bands of constant amplitude. A first band appears between 30 and 40 GHz with a span given by twice the chirp range. It is followed by a second band between 60 and 80 GHz. The third spectral band merges with the fourth one. Using an optical bandpass filter with sharp edges and with a full spectral width of 170 GHz (mixed grey line in Fig. 1(a)), we isolate the central component and the two first pairs of sidebands. The processing of the spectral phase is here achieved by a few $\pi/2$ phase shifts that replace the continuous triangular phase profile initially used in [8, 11]. Only four phase-shifts are inserted (at $\pm 25$ and $\pm 55$ GHz, Fig. 1(b), solid black line). Experimentally, it can be achieved using programmable liquid-crystal modulators [8, 11, 15] or fiber Bragg gratings [16]. We will compare the temporal properties of the pulse train with the results achieved when a quadratic spectral phase typical of a dispersive element is inserted. Different levels of cumulated dispersion $D$ are tested: 24, 32 and 44 ps$^2$ (solid, dotted and dashed grey lines in Fig. 1(b)), corresponding to the optimum level of quadratic phase that has to be imprinted to observe the best compression for modulation at a constant frequency of 40, 35 and 30 GHz respectively. Finally, we have also tested the results obtained in the presence of some possible bandwidth limitations of the optoelectronic devices. For the sake of simplicity and for the qualitative discussion, we have considered here that the various limitations that affect $A_m$ can be taken into account as a first-order lowpass filter, with a cutoff frequency of 40 GHz.

## 3. Results and discussion

Examples of the intensity profiles obtained after processing under different conditions are summarized for different instants of the frequency scan in Fig. 2. We can first note that the various schemes lead to intensity profiles with very different levels of background. The insertion of $\pi/2$ phase shits leads to waveforms of high quality, with a close-to-Gaussian profile, a duty cycle (here defined as the ratio of the full-width at half maximum duration and the pulse-to-pulse delay) and a peak power that remain constant when the optoelectronic bandwidth limitations are ignored. Note that processing a restricted limited number of spectral sidebands does not impair the temporal profile compared to the ideal case where the full spectrum is processed [8]. This strongly contrasts with the quadratic spectral phase that only partly compensates for the initial sinusoidal phase and that consequently induces much higher and detrimental background. Moreover, the pulse shape may vary according to the pulse-to-pulse delay. Whereas a cumulated dispersion of 44 ps$^2$ induces an optimal compression for the highest pulse-to-pulse delay (obtained at $t = -T_S$), it is a value of $D = 24$ ps$^2$ that leads to the best profile for the shortest pulse-to-pulse delay (achieved at $t = T_S$). A more quantitative study of the impact of the frequency linear chirp of the modulation on the main pulse properties is reported in Fig. 3 and 4. As can be seen in Fig. 3, in the ideal case ($\pi/2$ phase shifts, no optoelectronic bandwidth limitation, solid black line), the temporal duration may vary between -13% and +16% with respect to its average value, i.e. between 6.2 and 8.2 ps with remarkably no change of the peak power. Additional simulations (in the ideal conditions, results not shown here) have stressed that we can achieve for higher frequency span, modulation of the duration between -16% and +23% with respect to the average value. When limits of the modulation are taken into account (mixed black line), slight variations of the peak power ($\pm$ 5%) become visible and the range of variations of the fwhm duration is compressed. The results are significantly changed when a quadratic spectral phase modulation is involved. In that case, we can first note that the variations of the peak power are more pronounced and can reach as much as $\pm$ 20 % in the case of $D$ = 44 ps$^2$. The level of cumulated dispersion will also influence the range of variation of the fwhm duration that can be tuned from less than 10 % of the average value to more than 25 %. The link between the peak power and duration can remain monotonic (such as in the case of $D$ = 24 or 44 ps$^2$) or become more complex as in the case of $D$ = 32 ps$^2$.

Finally, we report in Fig. 4 the influence of the pulse-to-pulse delay on three parameters of the waveform. We can first note that the kurtosis excess that assesses the pulse shape [17] remains close to zero for the processing by $\pi$/2 phase shifts. That confirms that the waveform remains close to a Gaussian pulse, even in the presence of optoelectronics bandwidth limitations. On the contrary, when quadratic phase modulation is involved, the shape may significantly change as stressed by the large excursion of the kurtosis excess. Regarding the extinction ratio (defined here as the ratio of the peak power by the power at $\Delta T/2$ ), we note once again the poor performance achieved in the case of a quadratic spectral phase. With an extinction ratio always above 27 dB, the other scheme is by far much more promising. Finally, the duty cycle is found to be nearly constant to 0.25 in many configurations. However, when a quadratic spectral phase with a cumulated dispersion of 44 ps$^2$, is applied, we can observe that the duty-cycle can be nearly doubled.

## *4. Conclusions*

In order to conclude, we have extended the approach initially proposed in [8] and we have demonstrated that this scheme was also suitable with sweeping of the modulation frequency. Compared to the solutions based on an additional quadratic spectral phase, the insertion of a few discrete π/2 phase shifts enables a clear improvement of both the stability and the overall performances. Numerical simulations have revealed that the same Gaussian temporal shape can be maintained. However, in the presence of optoelectronic bandwidth limitations, some additional fluctuations of the peak power may appear.

The proposed architecture may find application in test and measurements where pulse trains with different pulse-to-pulse can be interesting and where testing different pulse durations can be wanted. With the progress of phase modulators [18], operation at extremely high frequencies up to 100 GHz can be foreseen, the optical spectral processing being not a limitation of the scheme. As the scheme is purely linear and does not require high-power, high signal to noise ratio can be expected.

***Acknowledgements:***

We acknowledge the support of the Institut Universitaire de France (IUF). We thank Ugo Andral, Julien Fatome, Alexandre Parriaux and Bertrand Kibler for fruitful discussions.

**Figure captions:**

Fig. 1 Optical spectrum. (a) Intensity profile of the phase modulated signal (black) and optical filter under use (grey dashed line). (b) Spectral phase profile applied to convert the phase modulation into temporal intensity modulation. Quadratic phase profiles corresponding to a cumulated dispersion $D$ of 24, 32 and 44 ps$^2$ are plotted with solid, dotted and dashed grey lines respectively whereas the spectral phase profile made of phase shifts of $\pi/2$ is shown with solid black line.

Fig. 2 Examples of the temporal intensity profiles obtained for different pulse-to-pulse delays $\Delta T$ of -$T_s$, 0 and $T_s$ ps are plotted in panels a, b and c respectively. The results obtained for a quadratic spectral phase profile (cumulated dispersion $D$ of 24, 32 and 44 ps$^2$ are plotted with solid, dotted and dashed grey lines, respectively) are compared with the results achieved using spectral phase shifts of $\pi/2$ without or with bandwidth limitation of the optical modulator included (solid and dashed black lines).

Fig. 3 Evolution of the peak power of the pulse as a function of the temporal fwhm duration of the pulses. The results obtained for a quadratic spectral phase profile (cumulated dispersion $D$ of 24, 32 and 44 ps$^2$ are plotted with solid, dotted and dashed grey lines, respectively) are compared with the results achieved using spectral phase shifts of $\pi/2$ without or with bandwidth limitation of the optical modulator included (solid and dashed black lines).

Fig. 4 Evolution of the excess kurtosis, extinction ratio and duty cycle (panels a, b and c, respectively) according to the pulse-to-pulse delay $\Delta T$. The results obtained for a quadratic spectral phase profile (cumulated dispersion $D$ of 24, 32 and 44 ps$^2$ are plotted with solid, dotted and dashed grey lines, respectively) are compared with the results achieved using spectral phase shifts of $\pi/2$

without or with bandwidth limitation of the optical modulator included (solid and dashed black lines).

Figure 1

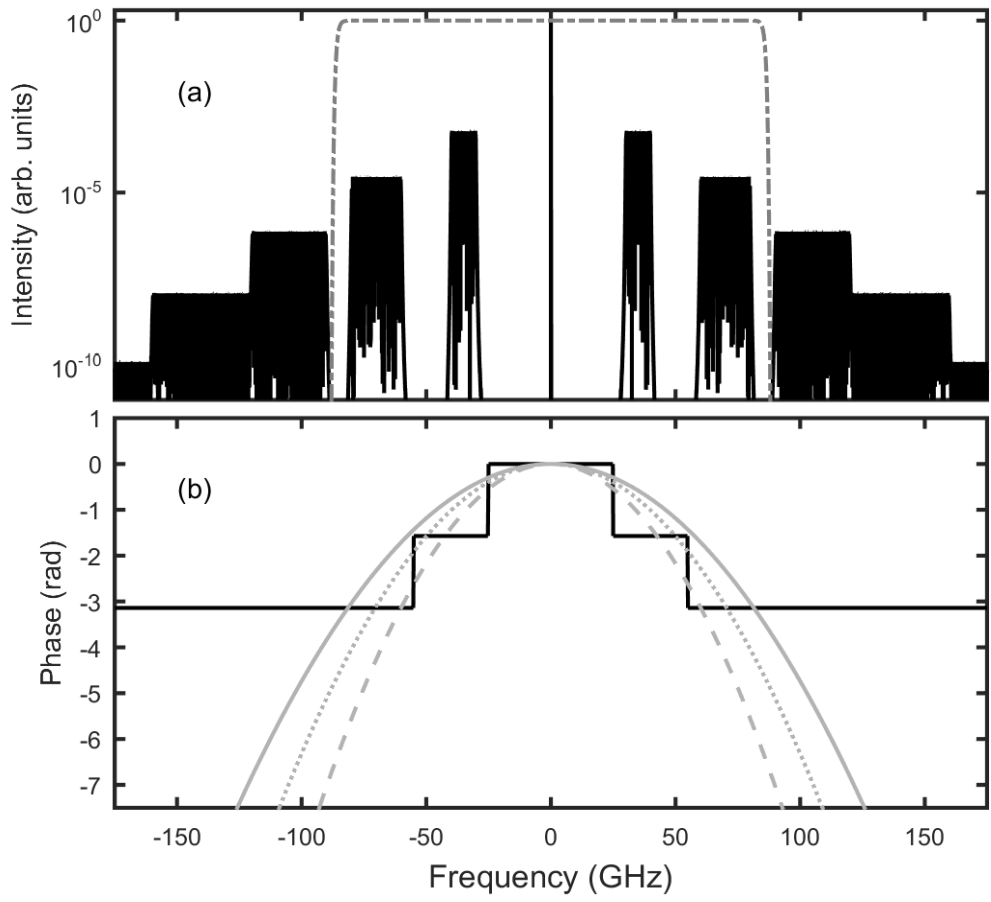

Figure 2

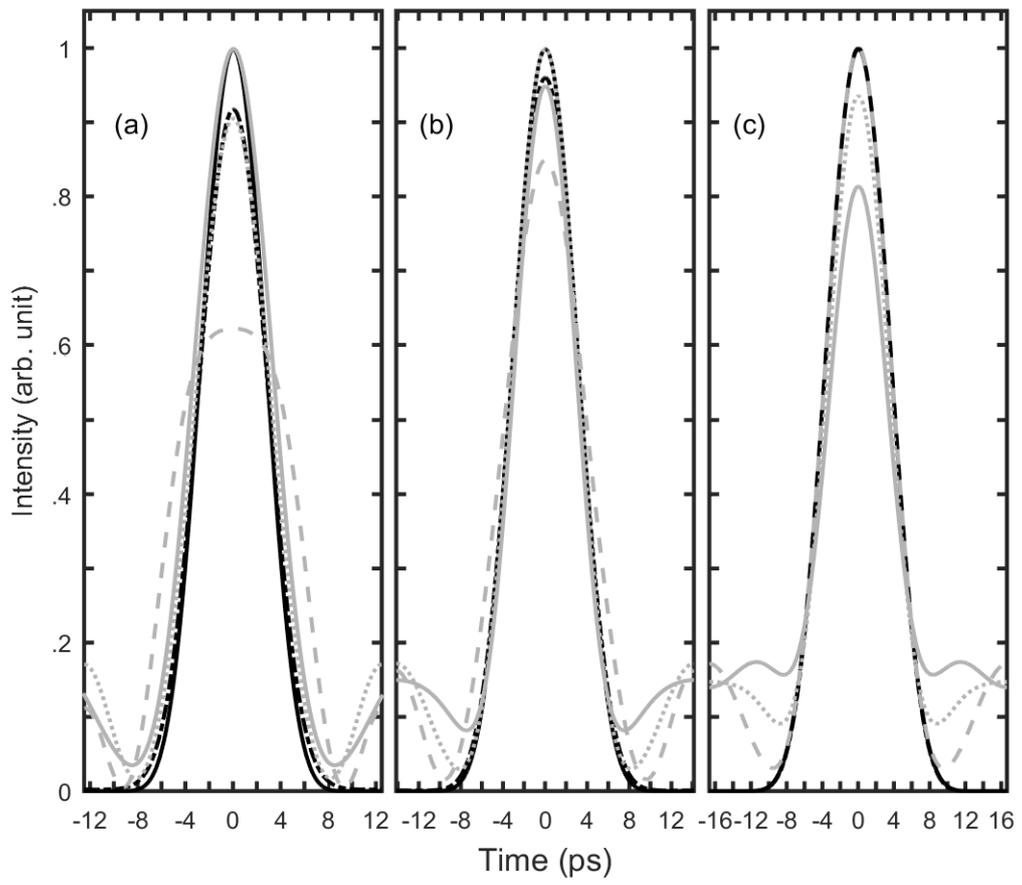

Figure 3

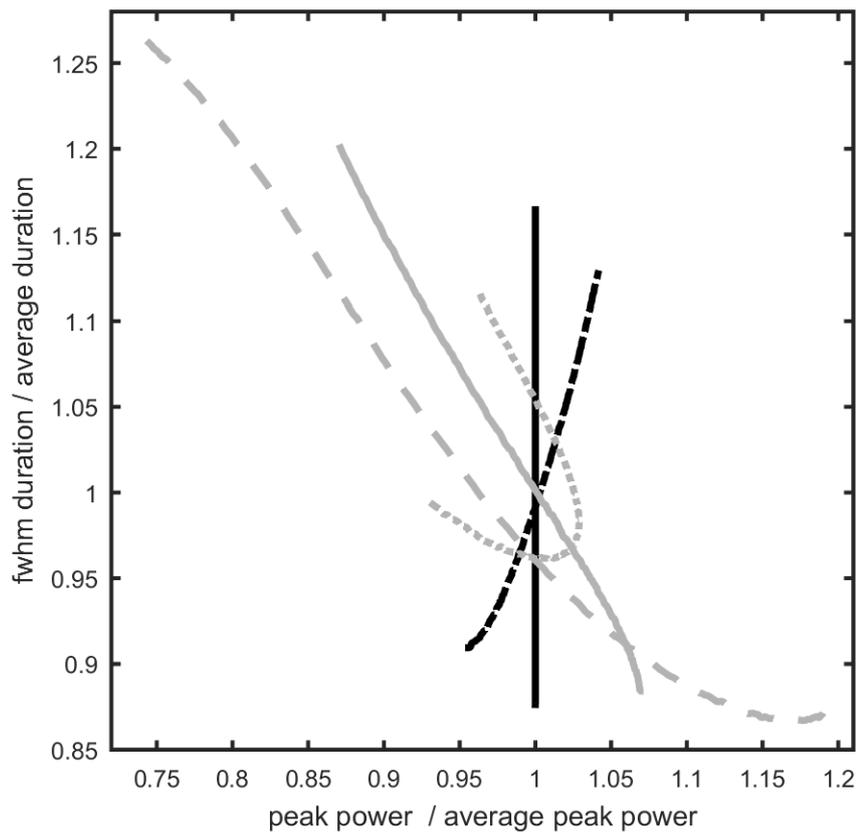

Figure 4

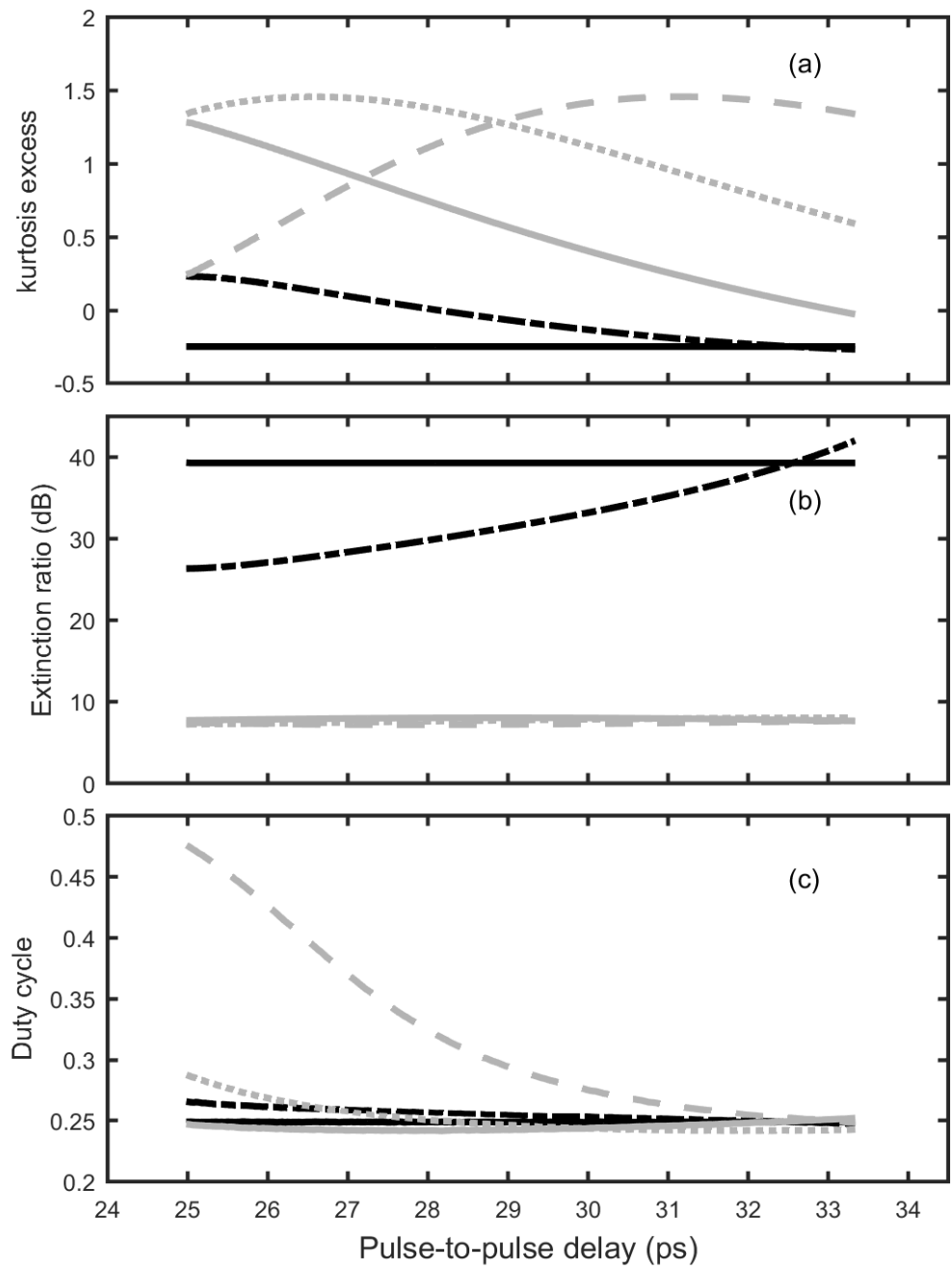